# Title:
Ghost Tomography

## Authors:
Andrew. M. Kingston*[1,2], Daniele Pelliccia[3], Alexander Rack[4], Margie P. Olbinado[4], Yin Cheng[4], Glenn R. Myers[1,2] and David M. Paganin[5]

## Affiliations:
1. Department of Applied Mathematics, Research School of Physics and Engineering, The Australian National University, Canberra, ACT 2601, Australia.
2. CTLab: National Laboratory for Micro-CT, Advanced Imaging Precinct, The Australian National University, Canberra, ACT 2601, Australia.
3. Instruments & Data Tools Pty Ltd, Victoria 3178, Australia.
4. European Synchrotron Radiation Facility, 38043 Grenoble, France.
5. School of Physics and Astronomy, Monash University, Victoria 3800, Australia.

* Correspondence to andrew.kingston@anu.edu.au.

## Abstract:

Ghost tomography using single-pixel detection extends the emerging field of ghost imaging to three dimensions, with the use of penetrating radiation. In this work, a series of spatially random x-ray intensity patterns is used to illuminate a specimen in various tomographic angular orientations with only the total transmitted intensity being recorded by a single-pixel camera (or bucket detector). The set of zero-dimensional intensity readings, combined with knowledge of the corresponding two-dimensional illuminating patterns and specimen orientations, is sufficient for three-dimensional reconstruction of the specimen. The experimental demonstration of ghost tomography is presented here using synchrotron hard x-rays. This result expands the scope of ghost imaging to encompass volumetric imaging (*i.e.*, tomography), of optically opaque objects using penetrating radiation. For hard x-rays, ghost tomography has the potential to decouple image quality from dose rate as well as image resolution from detector performance.

## One Sentence Summary:
We demonstrate ghost tomography with hard x-rays, thereby extending ghost imaging to a genuinely three-dimensional technique.

## Main Text:
Ghost imaging (GI) first emerged in the domain of visible-light optics *(1)*. The term arose from Einstein's description of quantum entanglement as 'spooky action at a distance' since initial realisations of the method utilised pairs of entangled photons. Classical implementations of GI have since been developed using pairs of correlated, coherent wavefields *(2)*. Very recently GI been achieved with atoms *(3)* and x-rays *(4-7)*. However, to date, none of the reported studies utilising penetrating radiation have attempted to map the interior structure of a genuinely three-dimensional (3D) sample. Ghost imaging clearly has the potential to achieve such tomographic reconstruction, constituting a natural extension of previously reported lower-dimensional ghost images. Here we report on the realization of ghost tomography using hard x-rays, whose penetrating power for optically opaque objects significantly extends both the applicability and utility of the technique.

Synthesizing images via the superposition of spatially random intensity maps is the essence of ghost imaging *(8-10)*. These maps may be generated through quantum processes such as shot noise or through classical means such as spatially random masks. A key feature of ghost imaging is that the ensemble of superposed spatially-random illuminating intensity maps is formed by photons (or other imaging quanta) that never pass through the sample. A weak copy of the illuminating field, which may be obtained *e.g.* using a beam splitter, does pass through the object but only the total number of transmitted quanta is measured by a single-pixel detector in a so-called 'bucket signal'.

Since no imaging quanta that pass through the sample are ever registered by a position-sensitive detector, ghost-imaging resolution is independent of the bucket detector. This is an important distinction between ghost tomography, and computed tomography: in computed tomography, 3D volume resolution is dictated by the pixel size of the detector; in ghost tomography, the correct 3D discretisation must be *found* based on analysis of the ensemble of illuminating spatially random fields. In two-dimensional (2D) ghost imaging applications, the parallelised intensity-intensity cross-correlation between the bucket and any one pixel of the random reference maps is used to compute the ghost image *(8,9)*. In what follows, we show that simply combining this method with tomography can be insufficient for 3D imaging and we present new reconstruction schemes that give superior results.

A schematic of our experimental setup for x-ray ghost tomography is shown in Fig. 1. Illumination of a spatially-random 1mm-thick Ni foam with normally-incident 26keV hard x-rays from a synchrotron, created spatially-random intensity illumination patterns such as that shown in Fig. 2A. An ensemble of such speckle patterns was obtained via transverse displacement of the foam over a 2D square grid with a step-size of 400µm. The sample for our x-ray ghost tomography experiment was an Al cylinder with diameter 5.60mm, into which were drilled two cylindrical holes with respective diameters of 1.98mm and 1.50mm (Fig. 2B). This sample was secured to a rotation stage, and illuminated with attenuated copies of the spatially-random intensity maps, obtained by using a 220 Laue x-ray reflection from a (001) Si-wafer beam-splitter.

Approximately 2000 random-illumination intensity maps were obtained, forming a linearly-independent mathematical basis *(10)* for the 2D tomographic ghost projections. As ghost-imaging spatial resolution *(11)* cannot be determined based on pixel size, we have used Fourier ring correlation to estimate the resolution of our imaging system as approximately 100µm (see supplementary material). FRC yields a best-case limit estimate for 2D spatial resolution. This is quite distinct from the point-spread function (PSF) of the 2D imaging system (Fig. 2C) that is calculated as the normalised auto-covariance of the set of illuminating spatially-random intensity fields *(11,12)*; this PSF estimates the resolution of conventional ghost-imaging by cross correlation *(8,9)* - see Supplementary Material for further detail. These estimates of resolution allow us to select an appropriate discretisation for our 3D reconstructed image.

For tomographic imaging, we repeated the 2000 random-illumination intensity maps for each of N=14 projection angles. Due to some instability of the beamline vacuum, the x-ray beam dropped out at random time intervals during data acquisition. Unlike conventional CT, ghost imaging is insensitive to such random signal-dropouts because it utilizes intensity-intensity *correlations*. Further, the object rotation angles were chosen using a quasi-random (or low-discrepancy) additive recurrence sequence of angles, Θ, with step size equal to ΔΘ = π(φ-1) radians, where φ=(1+5$^{1/2}$)/2 is the Golden ratio. This equates to ΔΘ = 111.25° and can be achieved equivalently with an angular step size for the object rotations of 180°-111.25°=68.75°. Quasi-random sequences appear to be random locally but are highly ordered globally; hence at any time the experiment is ceased, the angle set acquired will be an approximately uniform sampling of [0,π) radians.

Each detected image was registered onto an indirect detector, consisting of a scintillator screen, lens system and a 2560 x 2160 pixel pco.edge 5.5 (PCO AG, Germany) sCMOS-based camera with pixel-pitch of 6.5µm, and binned down to the resolution that was determined via Fourier ring correlation (FRC). Each object-free 2D reference-illumination beam was paired with a bucket-beam image containing the object (*e.g.*, blue box in Fig. 2D corresponding with yellow box in Fig. 2A). The total signal in the blue-box region was summed to give the bucket signal $B_{j,\Theta}$ corresponding to the j$^{th}$ realization of the spatially random illuminating pattern, at sample rotation angle Θ. The spatially-random intensity pattern $I_j(x,y)$ illuminating this same region corresponds to the spatially-resolved intensity map of the beam that did *not* pass through the object, where (x, y) are Cartesian coordinates in the detector plane.

In 2D ghost imaging, the cross-correlation ghost-imaging formula *(8,9)* is used to estimate the 2D intensity transmission function T(x,y;Θ) for a given fixed object rotation Θ as the ensemble average of $I_j(x,y)(B_{j,\Theta} - B_{av,\Theta})$ over j, where $B_{av,\Theta}$ is the average bucket signal for a given object orientation. Subsequent reconstruction of the 3D attenuation function using conventional tomography algorithms showed that the cross-correlation ghost-imaging formula is inadequate for 3D imaging, (*e.g.*, see Fig. S9). *A posteriori* information of the sample must be leveraged in order to produce something meaningful; to achieve this, we employed iterative cross-correlation via the Landweber algorithm coupled with smoothness priors *(13)*. The relaxation parameter used was $\gamma = 0.01/(J_\Theta \sigma^2)$, where $J_\Theta$ is the number of measurement pairs at angle Θ and $\sigma^2$ is the variance of the spatially-random speckle patterns. Such a 2D reconstruction was performed for each of the 14 pseudo-random projection angles Θ. Applying conventional tomographic reconstruction techniques to the resulting projection images produced a reasonable but very noisy tomogram (see Fig. S9).

In the above *two-step* reconstruction scheme (ghost reconstruction followed by tomography), each projection image is reconstructed separately from the others. This is not the optimal approach, as projections at different angles are obviously related. A better result can be achieved by *direct* reconstruction where one recovers the 3D volume directly from the bucket signals thus using all measured information simultaneously; the intermediate step of recovering the 2D x-ray ghost projection images can be removed. A gradient descent (or the Landweber) algorithm for direct iterative tomographic reconstruction from bucket signals has very recently been developed in Sec. V of the simulation-based study of Kingston *et al.* *(13)*. A smoothness prior and enforced positivity in attenuation coefficient were included here to improve the result. Vertical and horizontal slices through the resulting x-ray ghost-tomography reconstructions are shown in Fig. 3A and 3B respectively. These may be compared to the conventional, computed tomography reconstructions obtained in the same set of experiments, as given in Fig. 3C and 3D. The non-trivial pre-processing steps required to achieve the results in Fig. 3 are detailed in the supplementary material. The reconstructed sample densities, as obtained from the x-ray ghost tomograms, are quantitative. Using the XCOM (NIST) database *(14)* the attenuation per unit density of Al at 26keV is $1.65 cm^2/g$. The density of Al is $2.70 g/cm^3$ giving an expected linear attenuation coefficient of $4.455 cm^{-1}$. From the reconstructed x-ray ghost tomogram with 52µm pixel dimension (i.e. binned x8) the mean attenuation of the Al is measured as $4.80\ cm^{-1}$ and corresponds to the attenuation of Al at 25.3keV. This increase in attenuation is most likely due to inclusions of higher-Z metals to form the Al alloy, together with the difference in spectrum between the direct and diffracted beams.

The ability to achieve quantitative 3D imaging, in a ghost-imaging geometry where none of the photons passing through the object are ever detected with a position-sensitive camera, is remarkable. A key observation is the previously-mentioned impracticality of *two-step ghost tomography* achieved by simply combining 2D ghost imaging at each projection, with standard tomographic reconstruction concepts. Rather, we emphasize that *direct ghost tomography* was seen to be much more effective as iterative refinement occurs in a whole-of-dataset manner. We thereby reconstructed a 140 x 140 x 72 voxel ghost tomogram using approximately 26,000 bucket measurements spread over 14 sample-rotation orientations, equating to over 50 reconstructed voxels per bucket measurement. This efficiency was enabled by harnessing *a posteriori* assumptions (or enforcing 'priors'). Ghost tomography is particularly suited to such efficiencies, the further exploitation of which may aid in a long-term aim of reduced dose relative to conventional imaging.

From a broader perspective, our demonstration of ghost tomography shows how the ghost imaging approach is naturally able to decouple image quality from dose rate as well as image resolution from detector performance. This is a fundamental departure from conventional imaging paradigms. Ghost tomography affords the flexibility of independently varying a number of parameters, such as the illumination masks, exposure time, number of bucket readings and number of object orientation angles. As a consequence the resolution level can be optimised against dose rate in a way that takes into account prior knowledge about the sample. For instance illumination masks can be designed in a way to minimise dose to the sample (according to prior knowledge of it) while maintaining high resolution. This is not possible in direct imaging using a pixel array detector, which requires all pixels to be illuminated regardless of the object being imaged. Therefore, while it is important to compare the performance of ghost and conventional imaging - as done in this paper - it is crucial to recognize that ghost imaging is not just a different way of making images; a ghost imaging or tomography system can be designed to be 'adaptive', in the sense that it can be optimized for the features of the object being imaged. This may have great practical advantages when using ionizing radiation. For instance, it is not too far-fetched to imagine how ghost tomography with mask engineering could be used in future radiological practice. By using the available prior information, the dose could be spatially and angularly distributed to statistically match the object of interest (for instance the brain or the lungs) given the size, shape and density of these organs or body parts are well known.

In conclusion, we here report the first experimental demonstration of ghost tomography using x-rays. We show how ghost tomography is able to computationally measure the three-dimensional internal distribution of a sample by a set of zero-dimensional readings of the transmitted x-ray intensity from the sample. The task is accomplished by illuminating the sample with a known, varying set of two-dimensional x-ray patterns for each rotation angle of the sample. We discuss our strategies for data acquisition and processing, showing how direct tomographic reconstruction from bucket readings is much more effective than the two-step approach of tomographic inversion following ghost imaging reconstruction of individual projections. These results outline how the flexibility of engineering a ghost tomography measurement marks a radical departure from the conventional tomographic imaging paradigm, being able to make optimal use of the available information to maximize tomogram quality and minimize the radiation dose used.

**Acknowledgments:**
We acknowledge useful discussions with David Ceddia, Emilio Escauriza, Jean-Pierre Guigay and Timur Gureyev. **Funding:** AMK and GRM acknowledge the financial support of the Australian Research Council and FEI-Thermo Fisher Scientific through Linkage Project LP150101040. We thank the European Synchrotron Radiation Facility (ESRF) for granting beam-time on ID19, and for funding an extended stay by DMP. **Author contributions:** All authors contributed jointly to all aspects of this work. **Competing interests:** The authors declare no competing interests. **Data and materials availability:** All data needed to evaluate the conclusions in the paper are present in this paper and the supplementary materials.

**Supplementary Materials:**
Materials and Methods
- S1. Experiment design
- S2. Data pre-processing
- S3. Ghost imaging: recovering transmission images
- S4. Ghost tomography

Fig S1 – S10
References (15 - 18)

**Figures:**

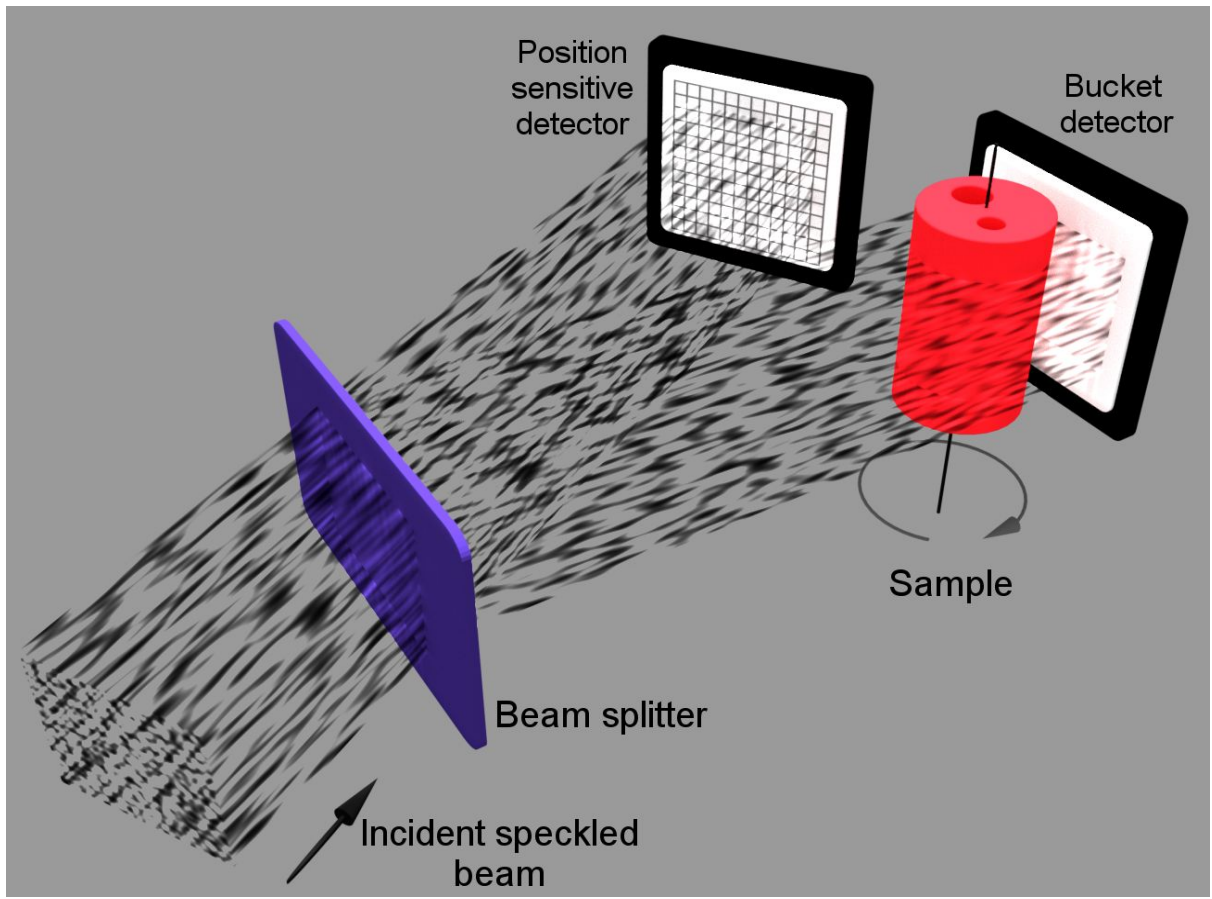

**Fig. 1.** Experimental setup for x-ray ghost tomography. Synchrotron x-rays from an undulator are passed through a spatially-random mask (not shown). The resulting random two-dimensional speckled beam is split into two copies by a crystal beam splitter working in a Laue diffraction condition. The diffracted beam, much weaker in intensity than the direct beam, is passed through the sample before being registered at the position-insensitive bucket detector. The direct beam, consisting of photons that never pass through the object, is measured over the position-sensitive detector. An ensemble of spatially-random illuminating patterns is created by transversely displacing the mask. Note that only the spatially-integrated signal (termed the 'bucket signal') for each bucket-beam measurement is utilized in the x-ray ghost tomography. The process is repeated for a variety of angular orientations Θ of the sample.

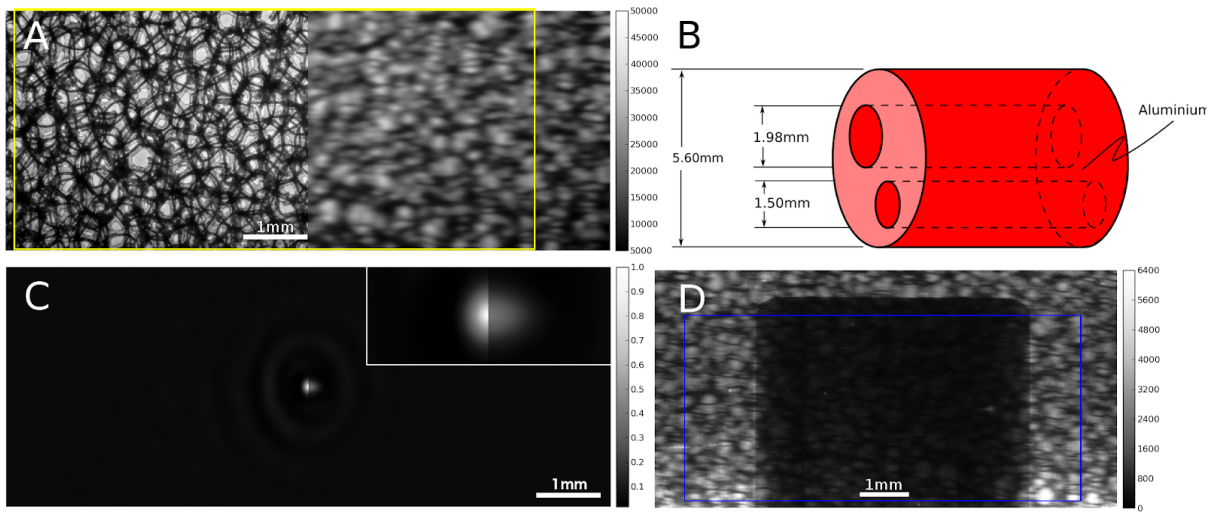

**Fig. 2.** A) Example spatially-random x-ray intensity illumination pattern; LHS: as measured, RHS: blurred to match motion artefacts in bucket image. Yellow box (coinciding with blue box in Fig. 2D) indicates region used for ghost-imaging/tomography. B) Schematic of Al phantom sample. C) Point-spread function (PSF) found as the normalised auto-covariance of the set of illuminating spatially-random fields; LHS: as measured, RHS: blurred to match motion artefacts in bucket image. Zoom x4 presented in top-left corner. D) Example bucket image with the blue box indicating the region over which the signal was accumulated to give the single-pixel bucket signal.

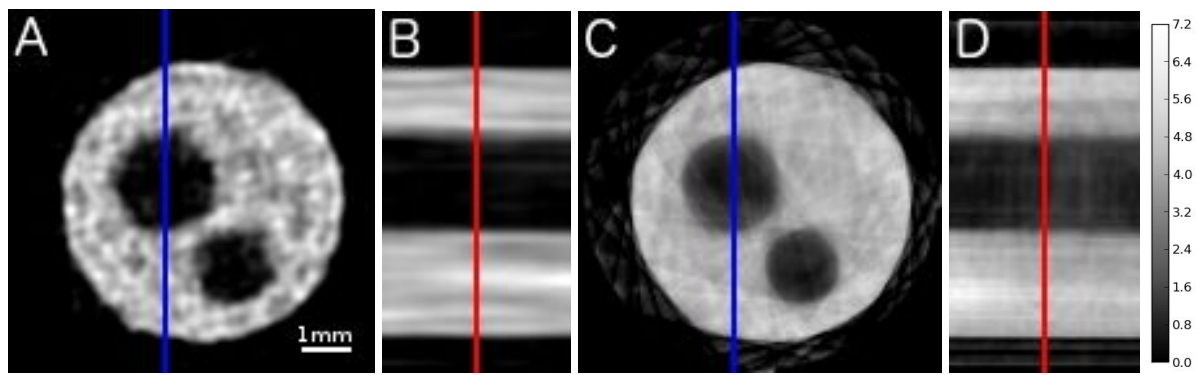

**Fig. 3.** Horizontal (A) and vertical (B) slices through the resulting x-ray ghost-tomography reconstructions with a voxel pitch of 48μm. Corresponding horizontal (C) and vertical (D) slices through standard tomography reconstructions obtained from the same set of experiments.

**Supplementary Material:**

**Materials and Methods**

**S1. Experiment details**

The experiment set-up is outlined in the main report. Speckle images were created by a Nickel (Ni) foam; this gave a good range of contrast with a mean transmission of 0.49 and a standard deviation of 0.243. Based on the dose-fractionation tomography simulations performed in *(13)*, we decided to use approximately 2000 speckle images per angular position. This was achieved by translating the foam over a 2D mesh of positions with 45 transverse steps of 0.4mm in both the vertical and horizontal directions; this gave a total of 2025 images per viewing angle.

The indirect detectors consisting of a LuAG:Ce (Ce-doped $Lu_3Al_5O_{12}$) single-crystal scintillator, 1x lens and 2560 x 2160 pixel sCMOS camera used for measurement have a pixel pitch of 6.5μm. The sCMOS recording the primary beam used 0.1s exposure while the sCMOS recording the beam reflected by the beam-splitter used 0.5s exposure. The speckle images had 8.696x greater intensity than the bucket images, indicating the silicon beam-splitter had a diffraction efficiency of 2.25% for the given incident x-ray spectrum. Each image was cropped to include only illuminated regions. The speckle images were cropped to 1500 x 600 pixels. Due to spatial distortions introduced by the beam-splitter crystal, the bucket image was cropped to 1448x751 pixels. Example speckle and bucket images are presented in Fig. 2 of the main report.

Several beam losses forced the experiment to stop/restart several times. Foreseeing this problem, we knew that we would be unsure how many viewing-angles of data we may be able to collect. Therefore we used the quasi-random additive recurrence sequence to order viewing angles as described in the main report. Collating measurements from the various experiment restarts into complete 'sets' of 2025 speckle/bucket image pairs resulted in 15 angles with a complete set of 2025 measurements at steps of 68.75° (modulo 360). Further analysis revealed that the X-ray beam was off for about 4000 consecutive measurements with the entire set of measurements for the 8th angle set (121.25°) missing and the 7th and 9th sets (52.5° and 10° respectively) only had about half of the 2025 speckle images each.

**S2. Data pre-processing**

A significant amount of pre-processing was performed prior to attempting ghost-imaging and ghost-tomography. Much of this could not have been done had an actual single-pixel bucket detector been used. However, if care is taken in pre-characterizing much of the effects described below, it will be feasible to use a bucket detector in practice. Here, we applied the following pre-processing steps to the speckle/bucket image pairs, and measured bucket values are estimated as the sum over all pixels in the bucket image. A multiplicative scale factor of 0.10984 is also applied to all speckle images to match the 'flatfield' regions of the corresponding bucket images.

### S2.1 Motion blur

The point-spread function (PSF) of our x-ray ghost-tomography experiment was calculated as the normalized auto-covariance of the ensemble of illuminating spatially random fields (Fig. 2C) *(11,12)*. This PSF has a full-width at half-maximum (FWHM) of 98µm. Observe however the significant horizontal blurring of the bucket images with respect to the speckle images due to the motion of the Ni foam during data acquisition; the movement of the Ni foam to the next position was triggered by the primary beam sCMOS which had 0.1s exposure *cf*. 0.5s exposure for the bucket beam CCD. It was predicted that the magnitude of motion artefacts would be acceptable being approximately that of the expected final ghost-imaging resolution. Comparison of Fourier power spectral density of the first speckle/bucket image pair (after applying a Hanning window) were used to find a blurring kernel that simulated the motion blur. The result (as presented in Fig. S1) is a Gaussian blurring kernel with $\sigma(x, y) = (71.5, 19.5)$µm. Post blur the FWHM of the PSF increases to 139µm vertically and 240µm horizontally.

Fourier ring correlation (FRC) *(15,16)* can provide an estimate of resolution by comparing the correlation of two independent measurements of the same object at various spatial frequencies, (*i.e.*, ring radii in Fourier space). Low correlation indicates a signal dominated by noise, and indicates the limit measurement resolution. Here we compared a 200 x 200 pixel (or 1.3 x 1.3 mm$^2$) image subset in the flatfield region beside the phantom, (see bottom row of Fig. S1), from the first speckle field at $\Theta = 0^0$ and $\Theta = 68.75^0$. The speckle images have full spatial resolution of 13µm (given a pixel dimension of 6.5µm). The bucket signal appears to have a reasonable resolution of approximately 25µm, however, we believe this is false resolution due to the correlation of crystal defects in the bucket images that are not affected by motion, (*e.g.*, the bright feature in the lower-right quadrant of the image in Fig. S1.) Correlating the speckle image with the bucket image confirms this indicating that the bucket image resolution is approximately 100µm and this corresponds well with the FRC analysis of the blurred speckle image with a raw speckle image.

### S2.2 Dark frames

The X-ray beam current was recorded in the header of the recorded speckle images. Speckle 'darkfield' images (when the current was 0.0) were observed to have a mean intensity per pixel of approximately 100 counts while a speckle 'flatfield' had a mean intensity per pixel of approximately 18,000 counts. All dark speckle/bucket image pairs were removed from the experimental data. An average of these pairs was used to estimate 'darkfield' images that were subtracted from all remaining images.

### S2.3 Registration

An approximate global alignment between the speckle and bucket images was found manually on the first speckle/bucket pair by matching the 'speckle-flatfield' part of the bucket images, (*i.e.*, beside the sample). An offset of $(x, y) = (455, 156)$µm and a scale of $(x, y) = (1.0382, 1.0256)$ was estimated. Resulting image pairs (including the simulated motion blurring of the speckle) are shown in Fig. S3. A per-image-pair refinement of this registration was then performed by maximising phase-correlation as described in Myers *et al. (17)*. The final dimensions of the registered full scale images became 1120 x 576 pixels.

*S2.4 Intensity normalisation*

The recorded synchrotron X-ray beam current exhibited a 'sawtooth' trend throughout the experiment with a variation of about 15% of the mean current. The variation is related to loss of electrons in the storage ring with time which is compensated by so-called refills appearing in equi-distant temporal intervals. The intensity of the speckle/bucket image pairs were normalized according to this beam current. Average speckle and bucket images were then computed at each angle (see *e.g.*, Fig. S4-T). A bright region may be observed at the top of the bucket image that corresponds to a dark region in the speckle image; the diffraction efficiency of the beam-splitter was higher in this region. Assuming a constant vertical profile of the flat-field regions over all average images, a scale was estimated for each image row to yield constant total counts per row. These average scale corrections were applied to all speckle and bucket images (see *e.g.*, Fig. S4-C).

*S2.5 Ring removal*

Ring artifacts were seen to arise due to non-idealities associated with the detector pixels, the associated x-ray scintillator, and the crystal beam-splitter. Affected pixels were identified using overall-average images, which should be smooth *a priori*, hence a median filtered image (using, in this case, a 5 x 17 pixel kernel) provided an estimate of the ideal average image. A per-pixel scale correction was identified from this and each measured image corrected accordingly. See Fig. S4-B for an example of the resulting image; *cf.* Fig. S4-C.

*S2.6 Find rotation-axis*

We determined average projected attenuation images at each angle as -log(B/S), where B is the average bucket image and S is the average speckle image. The horizontal position of the projected rotation axis was found by performing tomographic reconstruction by filtered back-projection (FBP) of these attenuation images with a range of horizontal shifts, h. An example reconstruction where h=0µm is given on the left of Fig. S5. As described in Kingston *et al. (18)*, the resulting tomogram with the sharpest reconstructed volume yields the optimal value for h. This was found to be h=234µm as depicted on the right of Fig. S5.

## S3. Ghost-imaging: recovering transmission images

*S3.1 Benchmark transmission images*

The transmission, B/S, and projected attenuation images, -log(B/S), can be estimated from the average of the speckle and bucket images at each angle. The images for Θ=0 are presented in Fig. S6. The transmission image in particular gives the objective function (or 'benchmark') for the performance of ghost image recovery in the following section; images of projected attenuation are required as the input for benchmark tomography in the next section.

*S3.2 Cross-correlation (XC) and iterative cross-correlation (IXC)*

Transmission images, $T(x,y;\Theta)$, at all 14 angles were recovered from the measured bucket values using (i) standard cross-correlation (XC) *(8, 9)*, (ii) iterative XC (IXC), and (iii) IXC with smoothness priors. Methods (ii) and (iii) were executed as described in Kingston *et al. (13)*. Some example images with the data binned 16x are presented in Fig. S7. In these cases, the iterative methods used 1120 iterations with $\gamma = 0.01/(J_\Theta \sigma^2)$, and the smoothness prior was executed as a Gaussian blur of the current estimate at each iteration with $\sigma(x, y) = (0.3,$

0.45)px. It can be observed that XC, or even IXC alone are insufficient; injecting priors in maximum *a-posteriori* methods, or compressed sensing is required to successfully extract the information present in the bucket measurements. In this case, injecting a smoothness prior significantly improved the ghost image.

**S4. Ghost-tomography**

We developed two methods of performing ghost tomographic reconstruction in Kingston *et al. (13)*: 1) standard tomography methods applied to recovered ghost-projection images, and 2) tomographic reconstruction directly from the bucket measurements. Here we have applied both methods for comparison. First, the objective (or benchmark) tomogram has been computed from the average projection images described in S3.1.

*S4.1 Benchmark tomography*

An iterative reconstruction technique is preferred given the limited number of viewing angles, 14, compared with 220 required to satisfy Nyquist angular sampling (when binned x8). Here we have used gradient descent iterative reconstruction (GD-IR) (Landweber iteration) of the average linearised projections to give a 'benchmark' tomogram for comparison with ghost tomography performance. The Landweber regularisation parameter used here was $\gamma = 0.5/(14N)$ where N is the tomogram dimension in voxels. We have presented the results for various resolutions, (binned 8x, 16x, and 32x), using 2N iterations of GD in each case.

*S4.2 Two-step tomographic reconstruction: tomography from ghost-projections*

As described in the main report, applying standard ghost-imaging techniques (cross-correlation (XC)) combined with standard tomography techniques (filtered back-projection (FBP)) is insufficient to yield acceptable reconstructions (see Fig. S9-L). Even using more sophisticated methods such as IXC to recover projection images followed by GD-IR to compute the tomogram produces a poor result (see Fig. S9-C). Here 3N iterations were used with a Landweber regularisation parameter of $\gamma = 0.2/(14N)$. Promising, but very noisy, results can be achieved by GD-IR applied to projection images recovered with IXC that include a smoothness prior (see Fig. S9-R). Here 2N iterations were used with a Landweber regularisation parameter of $\gamma = 0.25/(14N)$.

*S4.3 Direct tomographic reconstruction: tomography from bucket measurements*

Here we consider direct tomographic reconstruction from bucket values, *i.e.*, ghost tomography. The success of the two-step method described above (S4.2) is limited since each ghost-projection image is recovered separately. Corrections made per iteration of IXC use only data per angle. A reconstruction method that produces a volume directly from the bucket measurements uses the entire set of measured data per iteration; although slower, this can produce a superior result. In addition to this, priors applied in volume space are typically more powerful. For example, enforcing sparsity in gradient space by minimising total-variation would be a useful prior in volume space that is not necessarily applicable in projection space.

The theory for direct ghost-tomography was developed in Sec. V of Kingston *et al. (13)* for both weakly absorbing objects and generalised to the non-weakly absorbing case. Upon implementation, modifications were required to take the logarithm of XC images where XC yields negative numbers. We replaced XC of (B - <u>B</u>) with XC of (B - <u>B</u> + 0.05 B*) where <u>B</u> and B* are the mean and standard deviation of bucket values, B.

Again, gradient descent (or Landweber iteration) was used for iterative reconstruction. We used a Landweber regularisation parameter of 0.5/(14N) with 32N iterations. A smoothness prior was incorporated as well as enforced positivity to improve the result. Smoothness was reinforced per-iteration by blurring with a Gaussian kernel with $\sigma(x, y, z)$ = (0.25, 0.25, 0.5)px. Typically, iterative tomographic reconstruction proceeds from an empty (or zero) initial volume; we observed in this case that gradient-descent x-ray ghost tomography became trapped in local minima when using such a starting point. This is most likely due to the highly under-constrained nature of the problem; further research is required to determine if this is characteristic of iterative ghost-tomographic reconstruction. To overcome the problem we adopted a multi-scale approach where the initial seed for the iterative reconstruction (IR) at each scale was the prolongation of the output from the IR at the previous scale. At the coarsest scale, (namely 32x binning or 208μm voxel dimension), a zero initial estimate was used. The solution at each scale is presented in Fig. S10. Future work could involve a multi-grid solution, and consideration of more or different priors.

**Figures:**

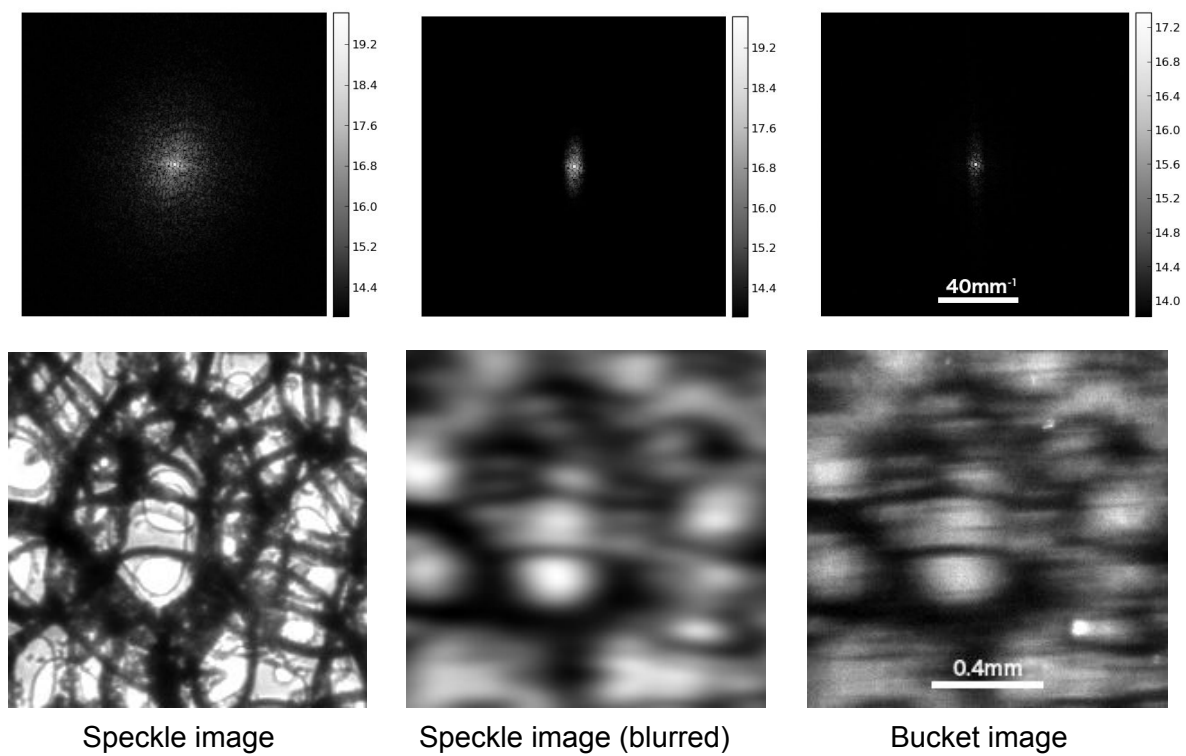

Speckle image        Speckle image (blurred)        Bucket image

Fig. S1. TOP: (L) Image of the Fourier power spectral density (FPSD, logarithm of the magnitude of the Fourier transform) for the example speckle image. (C) 2D FPSD of corresponding bucket image. (R) 2D FPSD of speckle image after Gaussian blurring with kernel $\sigma(x, y)$ = (71.5, 19.5)μm; This now matches the FPS of (C). BOTTOM: registered 200 x 200 pixel (or 1.3 x 1.3 mm$^2$) subset of corresponding images.

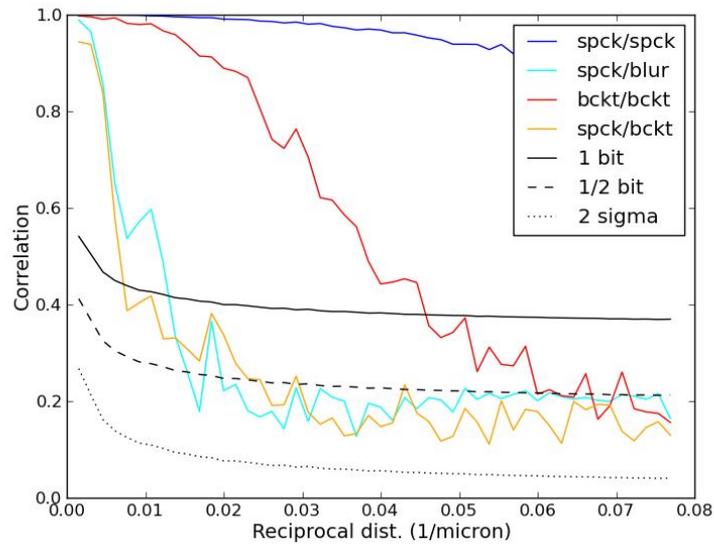

**Fig. S2.** Fourier ring correlation (FRC) results from 200x200 image subsets as exemplified in Fig. S1. Resolution for each image pair is determined as the reciprocal distance at which correlation drops below 1 bit. Image pairs include: spck/spck -- speckle images compared at $\Theta = 0^0$ and $\Theta = 68.75^0$; spck/blur -- speckle image at $\Theta = 0^0$ *cf.* Blurred speckle image at $\Theta = 68.75^0$; bckt/bckt -- bucket images compared at $\Theta = 0^0$ and $\Theta = 68.75^0$; spck/bckt -- speckle image at $\Theta = 0^0$ *cf.* bucket image at $\Theta = 0^0$.

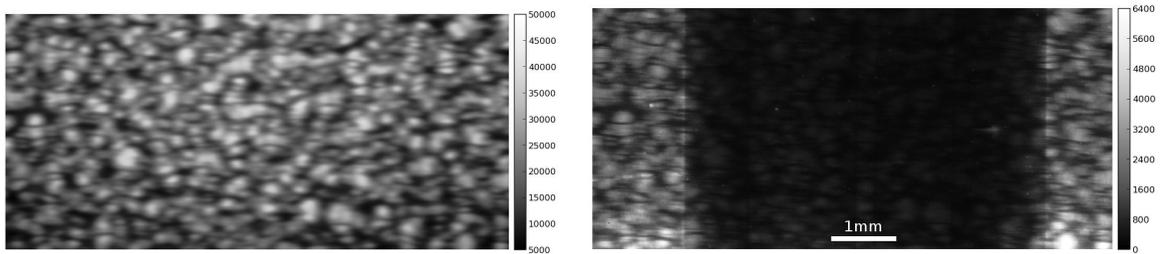

**Fig. S3.** (L) Example speckle image flipped vertically, magnified, translated, and cropped to match the speckle appearing in the (R) corresponding bucket image.

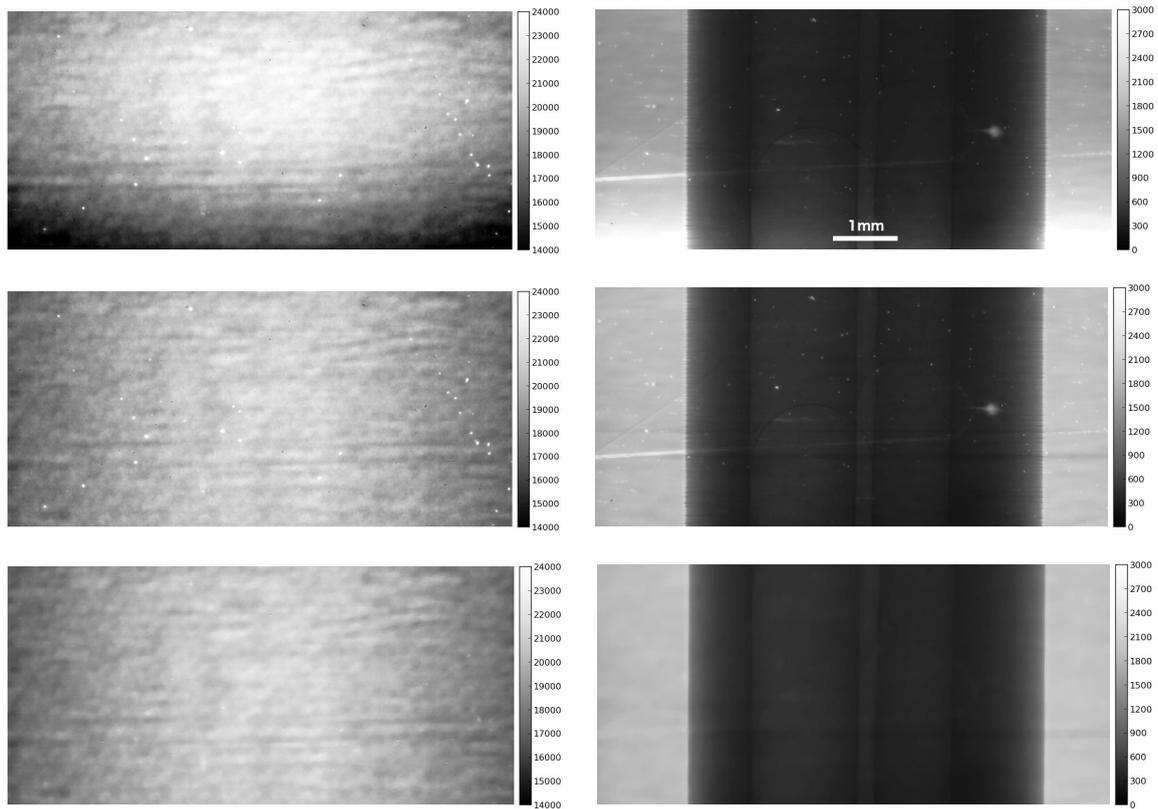

**Fig. S4.** Average images at Θ=0° of the (L) speckle and (R) bucket; (T) as measured, (C) after vertical flux variation was corrected, and (B) after potential origins of ring artifacts were suppressed.

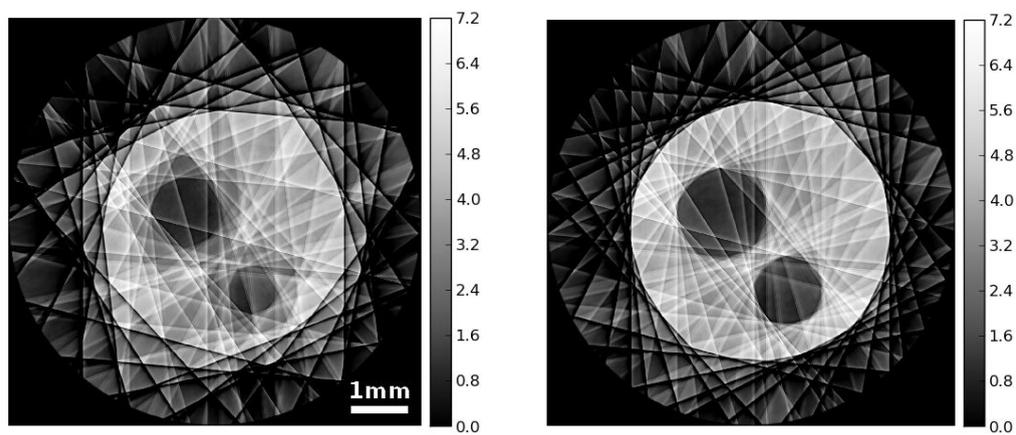

**Fig. S5.** Tomographic reconstruction by FBP of attenuation from the average images converted to projected attenuation, with various horizontal offsets of the rotation axis, h. (L) The result with h=0μm, *i.e.*, axis in the center of the images (R) the optimal result with h=234μm giving the sharpest tomogram.

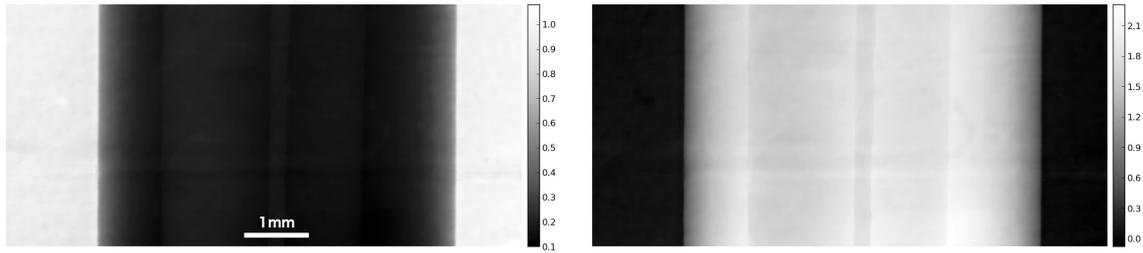

**Fig. S6.** (L) the average transmission image at Θ=0° (R) the corresponding linearised attenuation image.

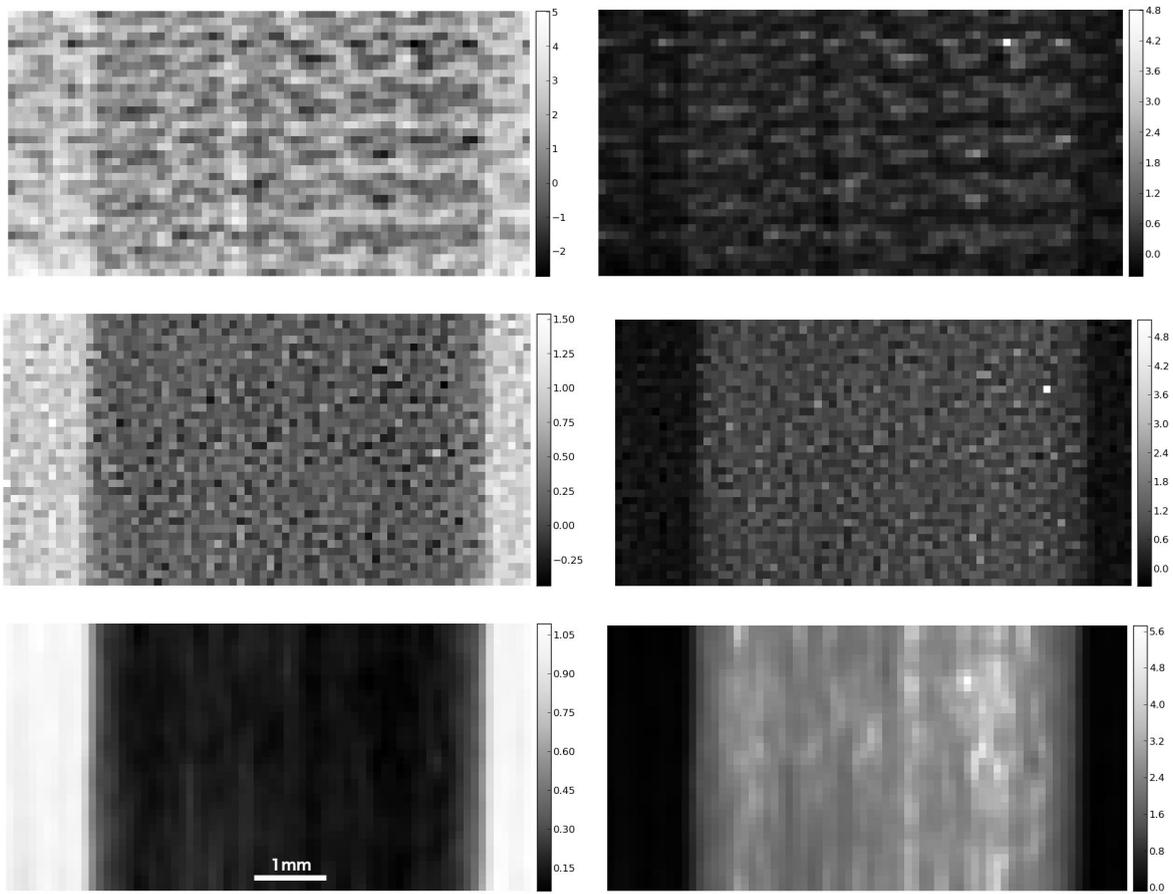

**Fig. S7.** (L) Transmission images, and (R) corresponding projected attenuation images, recovered from bucket measurements using (T) cross-correlation (XC), (C) 1120 iterations of iterative XC (IXC), and (B) 1120 iterations of IXC with a smoothness prior.

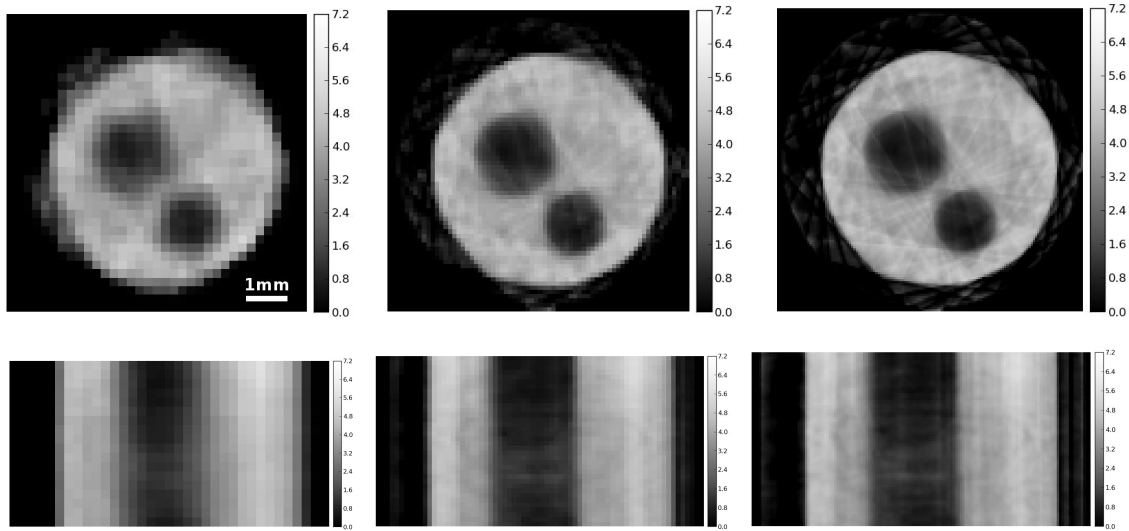

**Fig. S8.** (T) central xy-slices, and (B) yz-slices intersecting the large hole in the Al phantom, through tomograms resulting from GD-IR applied to average projected attenuation images. (L) Binned by 32, 2N = 70 iterations, (C) Binned by 16, using 2N = 140 iterations, (R) Binned by 8, using 2N = 280 iterations. All reconstructions used a 'zero' initial estimate with enforced positivity and a smoothness prior.

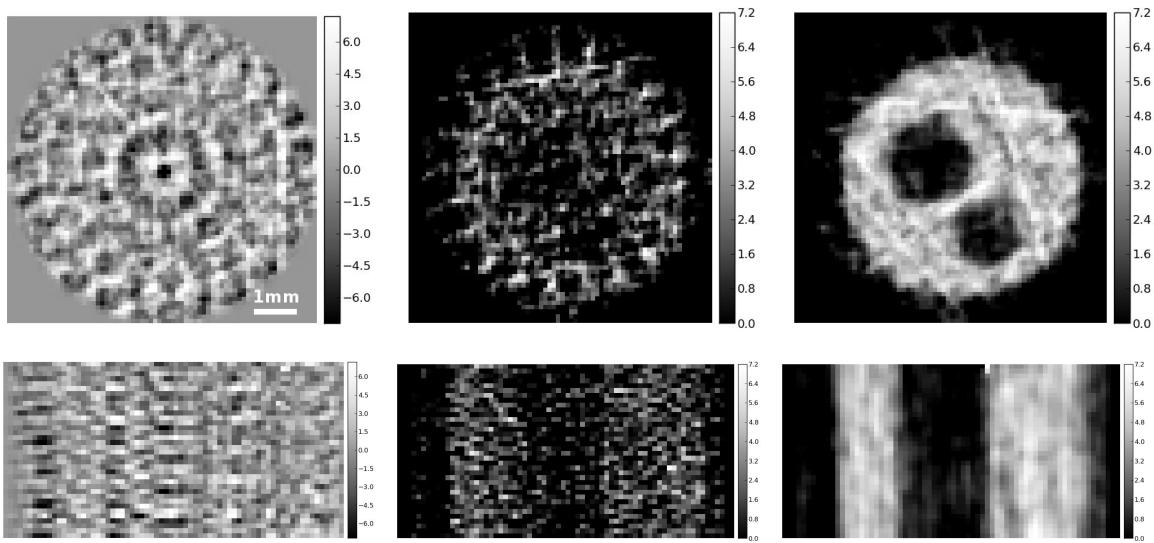

**Fig. S9.** Slices through tomography from ghost-images (binned 16x). (T) central xy-slices, and (B) yz-slices intersecting the large hole in the Al phantom. Reconstruction by (L) filtered back-projection (FBP) from XC projection images in Fig. S7-L, (C) 210 iterations of GD from the IXC projection images in Fig. S7-C, (R) 140 iterations of GD from projection images recovered through IXC with a smoothness prior (Fig. S7-R).

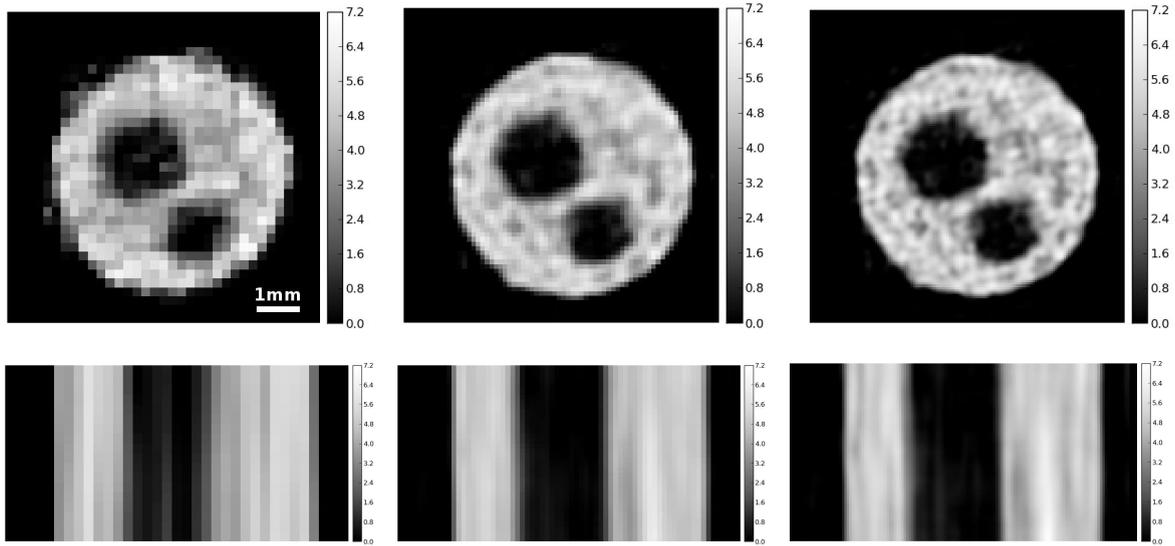

**Fig. S10.** (T) Central xy-slices, and (B) yz-slices intersecting the large hole in the Al phantom, through tomograms resulting from GD-IR with a smoothness prior in the volume applied directly to measured bucket values. (L) Binned by 32, with 'zero' initial estimate using 1120 iterations, (C) Binned by 16, with upscaled result of (L) as initial estimate using 2240 iterations, (R) Binned by 8, with upscaled result of (C) as initial estimate using 4480 iterations. All reconstructions enforced positivity and a smoothness prior.